\documentclass[11pt]{JHEP3}
\usepackage{amsfonts}
\usepackage{amssymb}
\usepackage{amsthm}
\usepackage{graphicx}
\usepackage{graphics}
\usepackage{multirow}

\usepackage{epsfig}
\usepackage{dcolumn}
\usepackage{bm}

\def\ie{{\it i.e.  }}
\def\eg{{\it e.g.  }}

\newcommand{\beq}{\begin{equation}}
\newcommand{\eeq}{\end{equation}}
\newcommand{\bea}{\begin{eqnarray}}
\newcommand{\eea}{\end{eqnarray}}
\newcommand{\phitc}{\frac{\phi_c}{T_c}}

\newcommand{\X}{{\rm X}}

\title{Electroweak baryogenesis window in non standard cosmologies}

\author{Gabriela Barenboim\\
	Departament de F\'{\i}sica Te\`orica and IFIC, Universitat de 
Val\`encia-CSIC, E-46100, Burjassot, Spain.\\
	E-mail: \email{gabriela.barenboim@uv.es}}

\author{Javier Rasero\\
	Departament de F\'{\i}sica Te\`orica and IFIC, Universitat de 
Val\`encia-CSIC, E-46100, Burjassot, Spain.\\
	E-mail: \email{javier.rasero@uv.es}}

\preprint{}	

\abstract{In this work we show that the new bounds on the Higgs mass are
more than difficult to reconcile with the strong constraints on the physical
parameters of the Standard Model and the Minimal Supersymmetric Standard
Model imposed by the preservation of the baryon asymmetry.
This bound  can be weakened by assuming a nonstandard cosmology at the time of the 
electroweak  phase transition, reverting back to standard cosmology by 
BBN time.  Two explicit examples are an early period of matter dominated 
expansion due to  a heavy
right handed neutrino (see-saw scale), or a nonstandard braneworld expansion.}


\begin{document}

\section{Introduction}
A wealth of cosmological observations over the past years have provided a deep knowledge on the thermal history of  the Universe 
since its first nanoseconds, up to today. 
Supernova candles show that the Universe is now accelerating as a consequence of 
an exotic particle  or more likely a cosmological constant with negative pressure. 
Measurements from the CMB tell us that our 
Universe is flat, isotropic and (almost) homogeneous and its physics can be accurately described by the Hot Big Bang Model 
and General Relativity. 

One of the corner-stones of the Hot Big Bang model is Big Bang Nucleosynthesis, the theory about the formation of light
elements (namely deuterium, helium, and lithium)  that were produced in the first few minutes after the Bang. The abundances of these 
lights elements depend on the density of protons and neutrons at the time of nucleosynthesis (as these were the only baryons 
around at this time) and provide a strong evidence for a necessity of a baryon asymmetry, an excess of nucleons over antinucleons.
Furthermore, the Universe seems to contain relatively few antibaryons. There is clear evidence that at least the local cluster
of galaxies is made of matter, and there is no plausible mechanism to separate matter from antimatter on such large scales.

Then one of the most challenging aspects of the interplay between particle physics and cosmology is to construct a compelling
and consistent theory that can explain the observed baryon asymmetry of the universe. The tiny difference between the number density
of baryons and antibaryons, of about $10^{-10}$ if normalized to the entropy density of the Universe.
In order to be able to generate such an asymmetry any theory must fulfil certain conditions. These conditions, called the Sakharov's conditions \cite{Sakharov:1967dj} 
establish the necessary ingredients for the production of a net baryon asymmetry, which are
\begin{enumerate}
 \item Non conservation of baryon number
\item Violation of C and CP symmetry
\item Departure from thermal equilibrium
\end{enumerate}

The need for the first two conditions is quite obvious. Regarding the third one the Universe must
have been out of thermal equilibrium in order to produce net bayon number, since the number of baryons
and antibaryons are equal in thermal equilibrium (if B violating processes do exist). It is also important to notice that {\bf all} known interactions
are in thermal equilibrium when the temperature of the Universe is between 100 GeV and $10^{12}$ GeV.
 
Many mechanisms for the production of the baryon asymmetry have been discussed for different periods
of the evolution of the early universe, which include GUT-baryogenesis, leptogensis, etc. Among all the proposals, 
the generation of the baryon asymmetry at electroweak scale is specially appealing since the electroweak scale
is the last instance in the evolution of the Universe in which the baryon asymmetry could have been produced within minimal frameworks. The Standard Model satisfies every Shakharov condition  and thus was considered that solely within this
framework baryogenesis could be explained. 

Firstly, baryon number violation occurs in the Standard Model through anomalous processes. Secondly, at low temperatures this anomalous baryon number violation only proceeds via tunnelling  which is exponentially suppressed. However, anomalous baryon number violation is rapid at high temperatures and the weak phase transition, if first order with supercooling, provides a natural way for the Universe to depart from equilibrium at weak scale temperatures. Electroweak phase transition can be then seen as bubbles of the broken phase which expand and end up filling the Universe. In this picture, local departure takes place in the vicinity of these expanding bubble walls. Lastly, $C$ and $CP$  are known to be violated by the electroweak interactions. So, in principle, all the required ingredients are there.

However, the standard model fails in almost every aspect. The CKM phase, the only source for $CP$ violation in the standard model, is extremely small to explain the observed baryon to entropy ratio. Another decisive check comes from the requirement that any net baryon asymmetry produced during the transition should survive until today. For an Universe whose expansion rate is slower than the anomalous baryon violating processes, thermal equilibrium would be recovered after the electroweak phase transition. Therefore, any asymmetry in baryon number created during the transition would be erased. In the broken phase, the rate of baryon number violation is exponentially suppresed by a factor $ {\cal O}\left(\phi /g T \right) $, where $\phi $ is the value of the order parameter and $g$ is the weak coupling constant. Thus, when demanding the baryon violating width to be smaller than Hubble rate, one finds
\beq\label{eq:sphb}
\frac{\phi(T_{ew})}{T_{ew}} \gtrsim 1 \, ,
\eeq   
where $T_{ew}$ stands for the temperature at which the electroweak phase transition is completed. Usually this temperature can be safely approximated to the critical temperature $T_c$ when both phases co-exist. The above condition constitutes the so called ``sphaleron bound",  and can give new information and constraints about the
$CP$ and Higgs sectors of the Standard Model. In particular, it has been
 shown that Higgs masses 
larger than 40 GeV can be ruled out by imposing that the baryon asymmetry of the Universe be generated
during the weak transition\cite{Kuzmin:1985mm,Arnold:1992rz}.

Nonetheless, the sphaleron bound (eq. \ref{eq:sphb}) presented above, assumes a particular thermal history
of the Universe,  one where during the electroweak 
phase transition the energy density of the universe was dominated by radiation. In section \ref{sec:sphb}, we will show that, under different thermal histories
of the Universe or different cosmologies, 
a less stringent condition can be obtained,  
permitting Higgs masses above the current experimental bounds. In section \ref{sec:sce}, we will analyse a scenario with a non standard 
thermal history during the electroweak phase transition which leads to a modified sphaleron bound condition, while in section 4 we relax this bound by modifying the underlying
cosmology. We will conclude in section 5.   

\section{Sphaleron Bound reviewed}
\label{sec:sphb}
The evolution of any baryon asymmetry  in comoving units during the electroweak phase transition can be written as 

\begin{equation}\label{eq:sph}
\frac{n_{\mbox{freeze}}}{n(t_B)} = \exp\left[-\int_{t_b}^\infty \ dt \ \widetilde{\Gamma}_{{\rm sph}(t)}\right] \, , 
\end{equation}
where $n_{\mbox{freeze}}$ is the baryon asymmetry which survives to partake of nuclesynthesis, $n(t_B)$ 
is the baryon asymmetry  at the beginning of the phase transition and  $t_b$ is the time at which the bubble nucleation proceeds, starting up the phase transition. 

The meaning of this equation is clear. The baryons created at the bubble walls are subject to decay after they
enter the broken phase, if the baryon number violating processes are not sufficiently suppressed. We should require then this
attenuation not to reduce the created asymmetry to less than that required for nuclesynthesis \ie
\beq
\int_{t_b}^\infty \ dt \ \widetilde{\Gamma}_{{\rm sph}(t)} = -\log\left(\frac{n_{\mbox{freeze}}}{n(t_B)}\right) \leq 1 \ .
\eeq

The sphaleron width is given by \cite{Carson:1990jm}
\begin{equation}
\widetilde{\Gamma}_{\rm sph}(t)= \alpha_n 6N_F^2 \ {\cal C} \ g \frac{\phi^7}{T^6} e^{-\frac{E_{\rm sph}}{T}} \ ,
\end{equation}
where $\alpha_n$ is a number of order one, whose precise value depends on the model and its corresponding set
of conserved charges and  $N_F$ is the number of fermion families. ${\cal C}$ is a temperature independent parameter
accounting for the degrees of freedom of the sphaleron and may be expressed in the following way
\beq
{\cal C} = \left(
\frac{\omega_-}{2\pi g\phi(T)}
{\cal N}_{\rm tr}{\cal N}_{\rm rot}
{\cal V}_{\rm rot}{\cal K}_{\rm sph}\right) \ .
\eeq
where $\omega_-$ is the frequency of the negative mode of the sphaleron, 
${\cal V}_{\rm rot} = 8 \pi^2 $, ${\cal N}_{\rm tr}{\cal N}_{\rm rot} \simeq 86 - 5 \ln (m_H^2/8 m_W^2)$ and 
${\cal K}_{\rm sph}= \{7.54, 5.64, 4.57, 3.89, 3.74\}$ for $m_H =\{0.4,0.5,0.6,0.8,1\}m_W$, and extrapolated for other 
values of $m_H$.

As the dominant contribution to the  integral (\ref{eq:sph}) comes from temperatures very close to $T_B$, it can be  
approximated to its value at this temperature. Such approximation slightly overestimates the dilution.

As $t\sim H^{-1}$, this yields the condition
\beq\label{eq:sphbound}
\widetilde{\Gamma}_{sph}(t_b) \leq H(t_b) \ .
\eeq
This equation shows what we pointed out before, the sphaleron rate processes must be slow enough, i.e out of thermal equilibrium, 
in order that any $(B + L)$ asymmetry won't be erased. This bound is usually stated as a lower bound on the sphaleron energy, or
as a lower bound on the ratio of the vev to the temperature at the critical temperature and can then be converted into a bound on
the parameters in a specific model.
    
Usually the literature shows this bound in the conventional cosmological scenario, that is, in a radiation dominated Universe. 
Within this scenario the expansion rate is given by 
\beq
H_{{\rm rad}}^2=\frac{4\pi^3}{45 M_{Pl}^ 2} g_* T^4\;.
\eeq

Inserting this expression in eq. \ref{eq:sphbound}, one finds that 
\beq\label{eq:sphhrad}
\frac{\phi_c}{T_c} \gtrsim \frac{1}{{\cal B}}\sqrt{\frac{4\pi}{\alpha_w}}\left(7\log\frac{\phi_c}{T_c} + \log W(T_c) - \log H_{{\rm rad}}\right)\;,
\eeq
where ${\cal B} =\{1.52,1.61,1..83,2.10\}$ for $m_H^2/m_W^2 \; \in \; \{0.008,0.08, 0.8,8\}$ and quadratically interpolated for intermediate values 
and  $W(T)= 6\alpha_n N_f^2 {\cal C} g T_c$. 
Solving this equation numerically gives 
\bea
\phitc \gtrsim 1\;.
\eea

Alternatively this bound can be restated  as a function/bound on different cosmological scenarios for which  the expansion  rate takes a different value. 
In such scenarios\cite{Joyce:1997fc}
\beq\label{eq:sphhhrad}
\frac{\phi_c}{T_c} \gtrsim \frac{1}{{\cal B}}\sqrt{\frac{4\pi}{\alpha_w}}\left(7\log\frac{\phi_c}{T_c} + \log W(T_c) - \log H_{{\rm rad} }\right) + \delta_\phitc\;,
\eeq
where 
\bea
\delta_\phitc = \frac{1}{{\cal B}}\sqrt{\frac{4\pi}{\alpha_w}}\log \frac{H}{H_{\rm rad}}\;.
\eea 

This new term has the effect of relaxing the sphaleron bound. This effect can be seen in figure \ref{fig:sphbH}, where the difference between the solutions given by eq. \ref{eq:sphhrad} and \ref{eq:sphhhrad}, {\it i.e,} $\Delta \left(\phitc\right) = \left.\phitc\right|_{H_{rad}} - \; \left.\phitc\right|_{H}$ is plotted for different values of $H$. 

In addition, it is clear that only drastic modifications, \ie  modifications where the energy density (and therefore the
expansion rate) is several orders of magnitude
larger than the one given in a radiation dominated scenario, can relax the bound in a sensible way.  
We are interested in studying whether such a modification to the sphaleron bound can be helpful to open up the allowed parameter
space for electroweak baryogenesis. 
To study this, let us review first how this bound is obtained in the Standard Model,  and what its implications are.

In the Standard electroweak theory  the effective potential at high temperatures reads as \cite{Cohen:1993nk}
\beq\label{eq:SMeffectivepot}
V(\phi,T) \approx \frac{M(T)^2}{2} - ET\phi^3 + \frac{\lambda_T}{4}\phi^4\;,
\eeq
where $M(T) , B$ and $\lambda_T$ are the temperature dependent effective mass, cubic term and quartic coupling respectively;
 given at the one-loop ring improved values
\bea
M(T)&=& \sqrt{A(T^2-T_0^2)}\, ,\nonumber\\
A &=& \frac{2m_W^2+m_Z^2+2m_t^2}{4 v^2}+
\frac{1}{2}\lambda_T\,,
\nonumber\\
E &=& \frac{2}{3}\left(\frac{1}{2\pi}\frac{2m_W^3+m_Z^3}{v^3}
+\frac{1}{4\pi}\left(3+3^{\frac{3}{2}}\right)
\lambda_T^{\frac{3}{2}}\,\right),
\nonumber\\
\lambda_T &=& \frac{m_H^2}{2v^2}-
\frac{3}{16\pi^2 v^4}\left(
2m_W^4\ln\frac{m_W^2}{a_B T^2}+
m_Z^4\ln\frac{m_Z^2}{a_B T^2}-
4m_t^4\ln\frac{m_t^2}{a_F T^2}
\right)\,,
\nonumber\\
T_0^2 &=& \frac{m_H^2+8\beta v^2}{2 A}\,,\qquad
\beta=\frac{3}{64\pi^2v^4}
\left(4m_t^4-2m_W^4-m_Z^4\right)\,.
\label{eq:effective potential II}
\eea
where $T_0$ is the temperature at which  the phase transition ends, 
$v = 246 {\rm GeV}$ is the usual Higgs vacuum expectation value at zero temperature, 
$a_B=(4\pi)^2{\rm e}^{-2\gamma_E}\simeq 50$, $a_F=(\pi)^2{\rm e}^{-2\gamma_E}\simeq 3.1$, and  
$\gamma_E$ is Euler's constant.

\FIGURE[t]{
 \centering
 \includegraphics[scale=0.7]{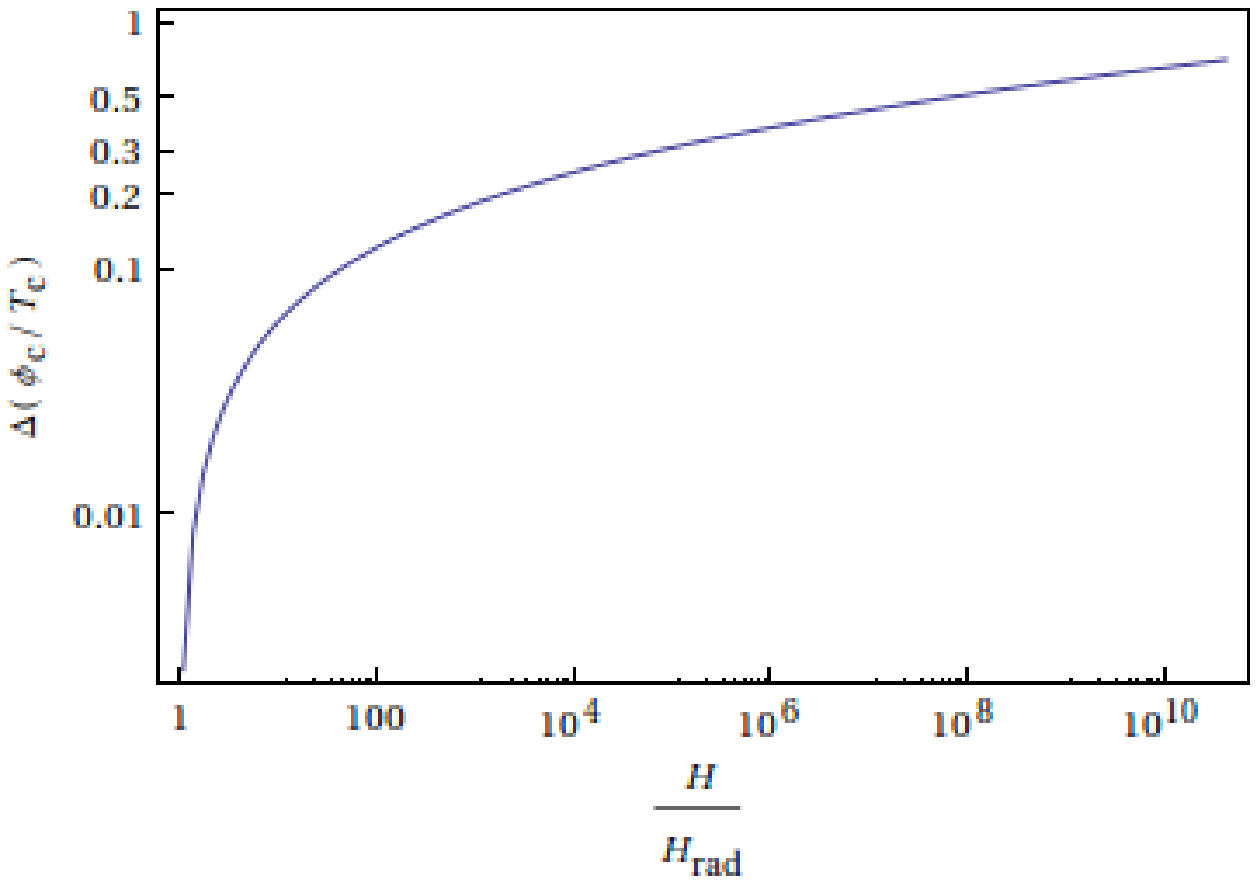}
\caption{Dependence of the relaxation of the sphaleron bound on the Hubble rate}
\label{fig:sphbH}
}

By minimizing the effective potential one finds that the ratio of the temperature dependent Higgs vacuum 
expectation value to the temperature  at a temperature at which a new degenerate 
minimum appears (the critical temperature) 
is given by
\beq
{\frac{\phi}{T}} = \frac{B+\sqrt{B^2 - 4\lambda_T A(1-\frac{T_0^2}{T^2})}}{2\lambda_T} \ .
\eeq

Using this result as a constraint on the model we can conclude that in order to have a sufficiently strong phase transition within the Standard Model 
the higgs mass should be smaller than $40 \ {\rm GeV}$, in
clear contradiction with current observations.
 This is why the Standard Model fails  to accommodate a mechanism to generate the baryon asymmetry during the electroweak phase transition. 
Nevertheless, we have already showed that the sphaleron bound could be weakened  by resorting to alternative thermal histories with significantly
different expansion rates at the electroweak phase transition. We have yet to see  whether the relaxation obtained can be large 
enough to allow current bounds on 
the Higgs masses. 
\FIGURE[t]{
 \centering
 \includegraphics[scale=0.8]{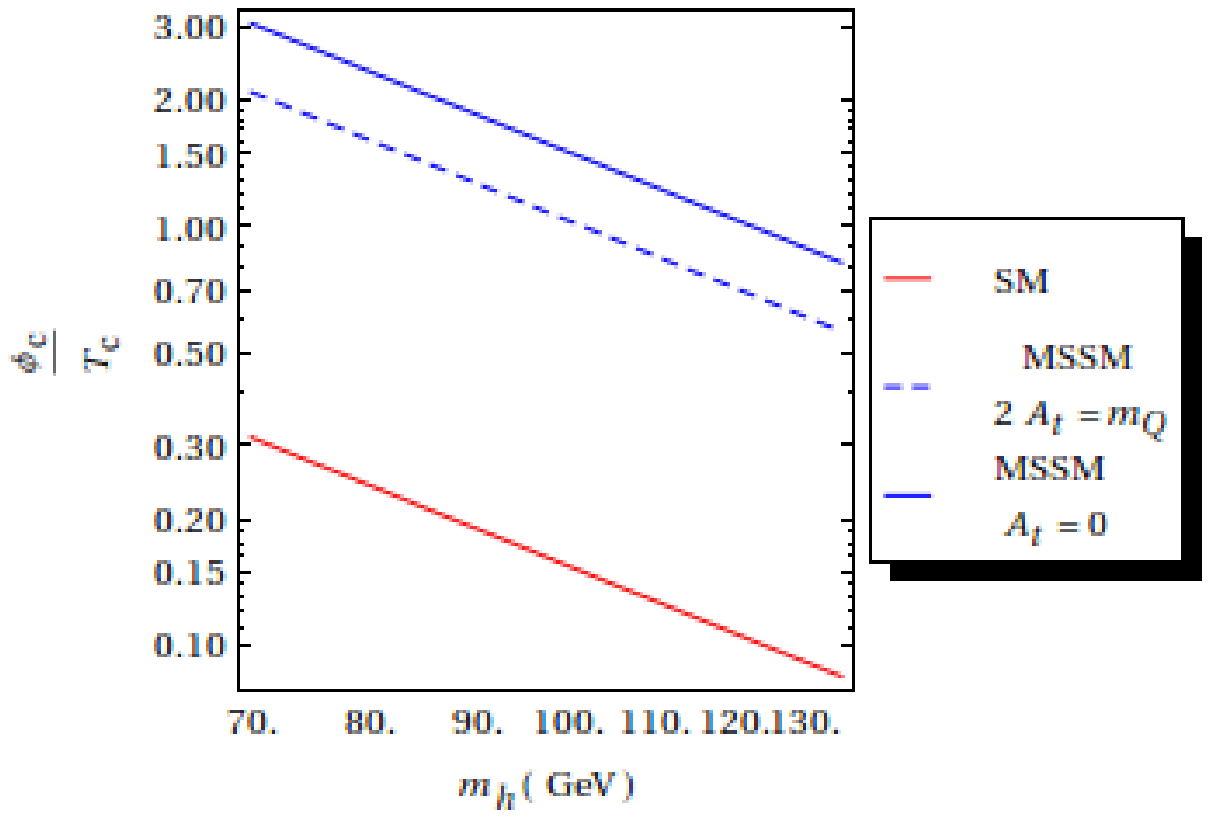}
\caption{Dependence of the ratio $\phi_c/T_c$ which controls the preservation of the baryon assymetry 
on the Higgs mass (treated here as a free parameter) for several models}
\label{fig:sphbSMvsMSSM}
}

There are in the literature plenty of extensions to the standard scenario with an enlarged matter sector where new effects appear and give rise to an enhancement 
of the strength of the phase transition. One of the most interesting extensions is supersymmetry and the so-called ``light stop scenario''\cite{CarDelLss}. 
In such scenario, stops are light enough, compared to the rest of superpartners, to affect the trilinear coupling to the higgs potential 
(finite corrections from heavy particles are highly suppressed). This effect impacts the ratio of the temperature dependent vev to the critical 
temperature in the following way
\beq
\left.\phitc\right|_{MSSM} = \left.\phitc\right|_{SM} + \frac{2 m_t^3}{\pi v m_h^2} \left(1- \frac{\tilde{A_t^2}}{m_Q^2}\right)^{\frac{3}{2}} \;,
\eeq 
with $$\tilde{A_t}=A_t-\mu/\tan\beta$$ the effective stop mixing parameter and $m_Q$ the soft supersymmetry breaking mass term for the stops. 
We can easily see that zero mixing makes the phase transition stronger so a parameter space  for this mixing close to zero is highly favoured. 
However, the mixing to the stops has also an important effect on the one loop corrections to the Higgs mass 
(in the decoupling limit, $M_A>> M_Z$, and a strong hierarchy in the stops spectrum)\cite{Espinosa:2001mm}

\bea
m_h^2=&& M_Z^2|\cos2\beta|^2 + \frac{3 m_t^4}{4\pi^2 
v^2}\left\{\log\frac{m_{\tilde{t}_r}^2 m_{\tilde{t}_l}^ 2}{m_t^4} \right.\nonumber\\ 
&& + \; \frac{A_t^2}{m_{\tilde{t}_l}^2}\left[2 \left(1 + \frac{m_{\tilde{t}_r^2}}{m_{\tilde{t}_l^2}}\right)-\frac{A_t^2}{m_{\tilde{t}_l}^2}\left(1 + 4\frac{m_{\tilde{t}_r^2}}{m_{\tilde{t}_l^2}}\right)\right]\log\frac{m_{\tilde{t}_l^2}}{m_{\tilde{t}_r^2}} \nonumber\\
&& + \; \left.2\frac{A_t^4}{m_{\tilde{t}_l}^4}\left(1 + 2\frac{m_{\tilde{t}_r^2}}{m_{\tilde{t}_l^2}}\right)\right\} 
\eea


Therefore, while non zero /strong  mixing enhances the Higgs mass, it does have the opposite effect on the strength of the electroweak phase transition. 
We can see this behaviour in the figures \ref{fig:sphbSMvsMSSM} and \ref{fig:HiggsmassMSSM}. So even in extensions to the standard scenario, a relaxation on the 
sphaleron bound would be welcome.

\FIGURE[t]{
 \centering
 \includegraphics[scale=0.8]{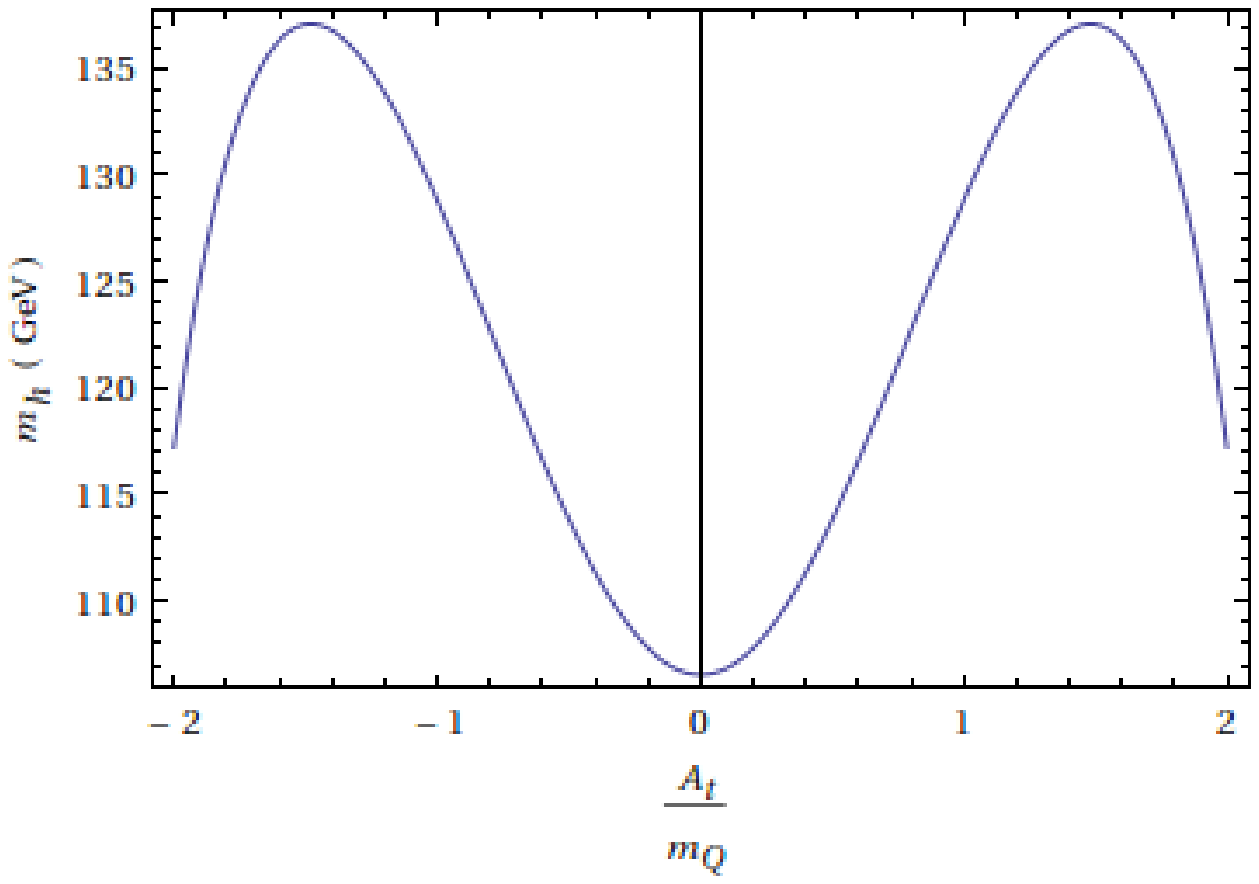}
\caption{Dependence of the Higgs Mass on the stop mixing. The stop mass $m_{\tilde{t}_l}$ is fixed and given approximately by the soft supersymmetry breaking mass term $m_Q$}
\label{fig:HiggsmassMSSM}
}

\section{First order phase transition in a matter dominated universe}
\label{sec:sce}

We have seen in the previous section that it is possible to make the electroweak phase transition strong enough to avoid the erasement of the asymmetry 
by sphalerons if  the value of the expansion rate at that scale is orders of magnitude larger than the expansion in the standard, radiation dominated, scenario. 
In the following, we will study a scenario where this condition can be naturally achieved. 

Because the Universe was extremely hot during its early stages, all kind of interesting particles (some yet to be discovered, some which hasn't 
even been postulated)  were present in significant amounts.  For $T \gg m$, the mass of the particles in question, their equilibrium abundance is,
to within numerical factors, equal to that of photons. When the temperature of the thermal bath drops below $m$, the equilibrium abundance 
of such particles is less than
that of photons and their  contribution to the total energy density becomes suppressed by a factor,
\beq
(m/T)^{5/2} \exp^{-m/T}\;,
\eeq
{\bf except} if one (or more) of such particles, which in the following we will
call $X$, drops out of equilibrium and its abundance freezes out 
(we are assuming that $X$ annihilation cross section is very suppressed). 
 In this case, the relic abundance
of $X$  relative to photons remains approximately constant and the contribution to the energy density of $X$ grows as $1/T$ as compared to that of photons.
It is obvious then, that eventually the energy density of $X$ will dominate that of the Universe. If the $X$ particle is  unstable (but long lived enough) 
and decays into 
relativistic particles which thermalise (releasing large amounts of entropy) the Universe will re-enter a radiation dominated era.
This will be the scenario we will focus on.

If we assume a flat Universe (as given by observations) the evolution equations for the different components of the Universe are given by
\bea
\label{eq:fulleqs}
\dot{\rho}_X &=& - 3 H \rho_X - \Gamma_X \rho_X \\[.1cm]
\dot{\rho}_r^{\mbox{\small{old}}} & = & - 4 H \rho_r^{\mbox{\small{old}}} \\[.1cm]
\dot{\rho}_r^{\mbox{\small{new}}} & = & - 4 H \rho_r^{\mbox{\small{new}}} + \Gamma_X \rho_X  \\[.1cm]
H^2  & = &\frac{8 \pi}{3 M_{Pl}^2} \left(\rho_X + \rho_r^{\mbox{\small{old}}} +
\rho_r^{\mbox{\small{new}}}\right)\;,
\label{eq:fulleqsb}
\eea
where $\rho_X$ is the energy density associated to the particle $X$, once it becomes nonrelativistic 
and $\Gamma_X $ its decay width,
$\rho_r^{\mbox{\small{old}}}$ is the energy density in radiation not
associated with $X$ decays, while $\rho_r^{\mbox{\small{new}}}$ is the one coming from $X$ decays \footnote{Evolution 
equations with tracking, \ie when the different components of the Universe chase each others abundance, can also produce
early periods of matter domination \cite{Barenboim:2005np}}.

\FIGURE[t]{
 \centering
 \includegraphics[scale=0.8]{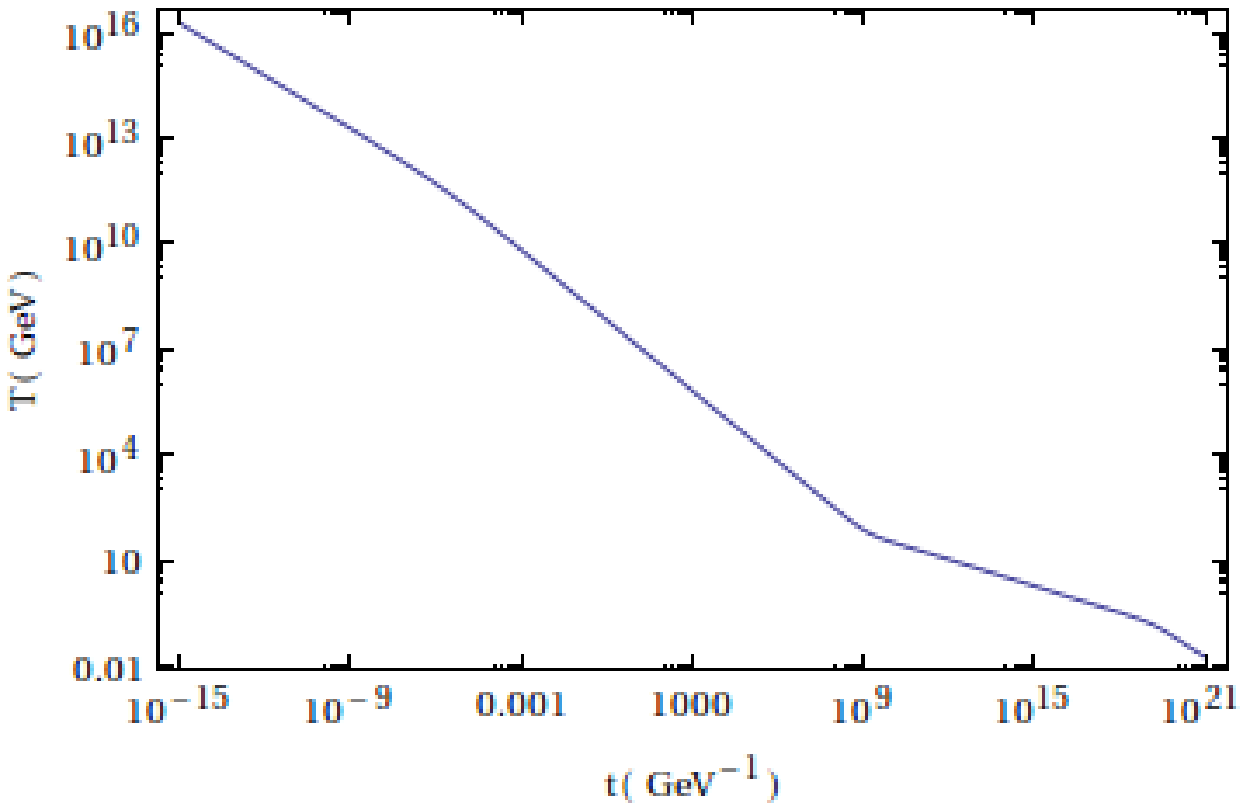}
\caption{Evolution of the Temperature in an Universe which goes through different epochs}
 \label{fig:temp1}
}

Contrary to the standard picture, the temperature of this universe, will have two sources
\bea
T(t)= \left( \frac{30}{\pi^2 g_*} \left(\rho_r^{\mbox{\small{old}}}(t) + \rho_r^{\mbox{\small{new}}}(t) \right) \right)^{1/4}
\eea
and therefore  its temperature profile, shown in figure \ref{fig:temp1}, will be significatly different from that of the standard case\cite{Scherrer:1984fd}.
We will start at temperatures larger than the mass of our particle $X$,  $M_X$, with a radiation dominated universe, where
$X$ is in thermal equilibrium. During this time the temperature scales like $t^{-1/2}$ (or $1/a$, being $a$ the scale factor).
Once the temperature drops to 
\beq
T_{\rm start} = \frac{4}{3} r M_X\,,
\eeq
with $r = g_X/2$ if $X$ is a boson and $r=3 g_X/8$ if it is a fermion, being
$g_X$ the total number of spin degrees of freedom of $X$,  we enter a matter dominated period. 
During the first part of this period, which comprises most of the matter dominated era, and although $X$ is decaying through an
exponential law
\bea
\rho_X \simeq \frac{2 \pi^2 g_*}{45} r M_X T^3  e^{-\Gamma_X t}\,,
\eea
the exponential factor does not affect in a significant way $X$ abundance, 
the radiation released by $X$ decays is negligible compared with that
not coming from $X$ decays,
and the temperature
falls as in a pure matter dominated period, \ie  $\; T \propto t^{-2/3} \propto 1/a $.

As $t$ approaches $1/\Gamma_X$, the new radiation starts to be comparable with the
old one. Thus, $X$ quickly dissapears into (new) radiation  and $T \propto t^{-1/4} \propto 1/a^{3/8}$. 
Once the age of Universe exceeds $1/\Gamma_X$, our matter dominated Universe 
turns into a radiation dominated
one and the temperature starts once more to track the scale factor $T \propto t^{-1/2} \propto 1/a$.
At this point
\beq\label{eq:tempdecay}
T_{{\rm end}} = 0.78 g_*^{-1/4}\sqrt{M_{Pl}\Gamma_X}\ . 
\eeq

\FIGURE[t]{
\centering
\includegraphics[scale=0.8]{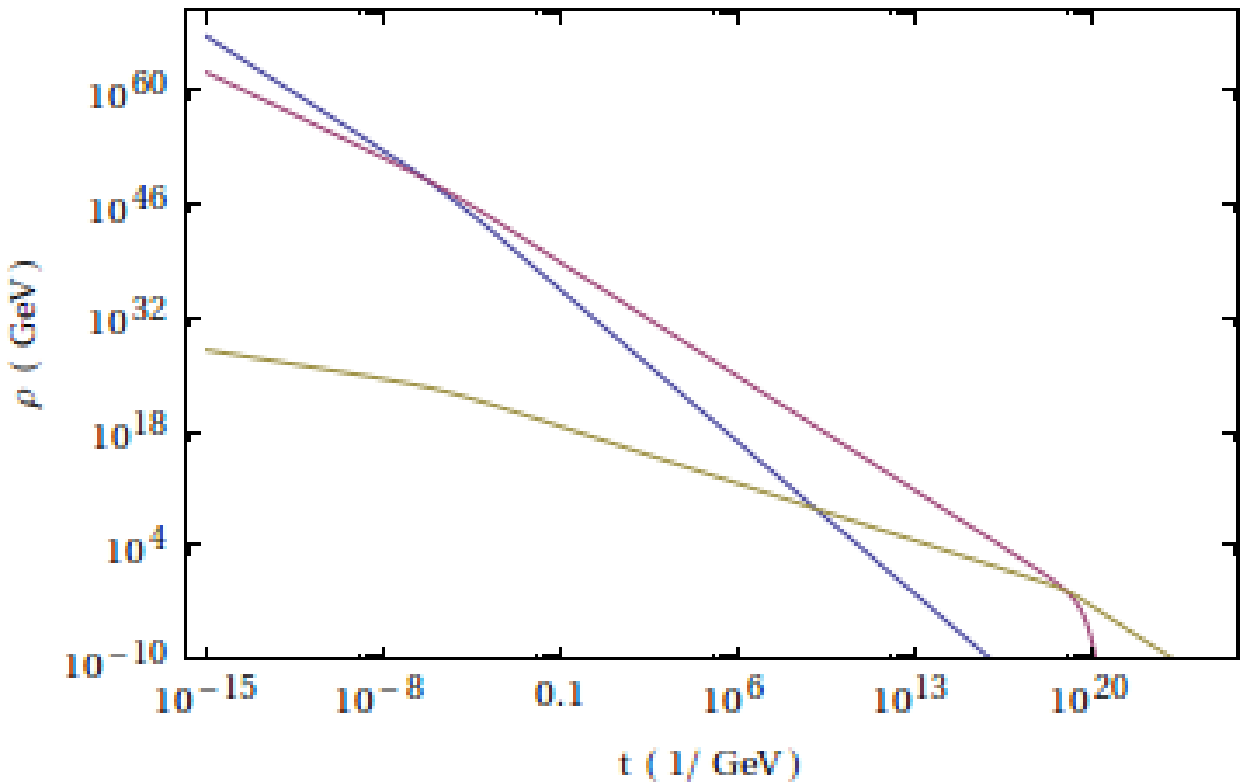}
\caption{Evolution for the different components of the Universe. Purple, blue and brown line correspond to $\rho_X , \rho_r^{\mbox{\small{old}}} ,\rho_r^{\mbox{\small{new}}}$ respectively. In this case, the decay width is $\Gamma_X = 10^{-19} \ {\rm GeV}^{-1}$  }
\label{fig:history1}
}

The described stages the Universe goes through are depicted in figure \ref{fig:history1}, for a particular choice of parameters. Since we want to recover the standard (radiation dominated) picture  before nucleosynthesis, we can derive a
lower bound on $\Gamma_X$ by requesting that the Universe must have left at the latest the matter dominated era shortly before BBN, which reads
\beq
\Gamma_\X \geq 2.0\cdot10^{-24}\sqrt{\frac{g_*}{200}} \ {\rm GeV}\;.
\eeq

As stated in the previous section an essential condition  for electroweak baryogenesis is that the sphaleron
transitions would be turned off after the phase transition so that no washing out of the asymmetry produced
during the transition occurs. This situation is achieved when the transition rate of the sphaleron
interactions is small as compared to the Hubble rate \ie, when these transitions are out of equilibrium.
In a radiation dominated universe, the Hubble parameter scales like $H \propto T^2$, however in a matter dominated one, during its first period, when the decays does not significantly reduce the
abundance of the $X$ particle
\beq
H_{{\rm MD}}^2 =\frac{16\pi^3 g_* r}{135 }  \left(\frac{M_{\rm X} T^3}{M_{Pl}^2}\right)\;.
\eeq
It is then straightforward to notice that large $M_{\rm X}$ masses can substantially change the value 
of the Hubble rate at electroweak scale and consequently affect the strength of the phase transition. 

In order to quantify this variation, we must distinguish between two cases: 

(i) The electroweak
phase transition temperature is reached when the temperature is 
essentially given by radiation not coming from $X$ decays, \ie 
\bea
\rho_r^{\mbox{\small{old}}}  \big\vert_{\mbox{\small{EW}}} 
\; \; \gg  \;\; \rho_r^{\mbox{\small{new}}}\bigm\vert_{\mbox{\small{EW}}}\ ,
\eea
 where in terms of the model parameters
\bea
\rho_r^{\mbox{\small{new}}} \simeq  0.221\;\Gamma_X \; M_{Pl} \; \sqrt{ g_* T_{\rm EW}^3 M_X} \;
\eea
implies that in this scenario we require
\bea
\Gamma_X \ll \frac{2.46 \times 10^{-13} }{\sqrt{r} }\sqrt{\frac{(\mbox{GeV})}{M_{\rm X}}}\;. 
\eea
In this case, the extra contribution to the sphaleron bound is given by
\beq
 \delta_\phitc = \frac{1}{{2 \cal B}}\sqrt{\frac{4\pi}{\alpha_w}}\log \frac{4 r}{3} \frac{M_X}{T_{\rm EW}}
\eeq
so that it gets modified as
\beq
\Delta\left(\phitc\right) \approx - \frac{\log \frac{4 r}{3} \frac{M_X}{T_c}}{ \frac{1}{{\cal B}}\sqrt{\frac{4\pi}{\alpha_w}} - 7 \frac{1}{\left(\phitc\right)}}\; .
\eeq
(ii) The temperature at the phase transition is set by the radiation coming from $X$ decays, \ie
\bea
\rho_r^{\mbox{\small{new}}} \bigm\vert_{\mbox{\small{EW}}} \; \; \gg  \;\; \rho_r^{\mbox{\small{old}}}\bigm\vert_{\mbox{\small{EW}}}
\eea
so that
\bea
\frac{\rho_X}{\rho_r^{\mbox{\small{new}}}} \simeq \frac{ g_*}{1.9 \times 10^{33}} 
\left(\frac{ T_{\rm EW}}{\Gamma_X} \right)^2
\eea
and the relaxation on the sphaleron bound reads
\beq
\Delta\left(\phitc\right) \approx - \frac{ \log \frac{ g_*}{1.9 \times 10^{33}} \left(\frac{ T_{\rm EW}}{\Gamma_X}\right)^2 }{ \frac{1}{{\cal B}}\sqrt{\frac{4\pi}{\alpha_w}} - 7 \frac{1}{\left(\phitc\right)}}\;.
\eeq

From these equations it is clear that although both schemes can significantly relax the sphaleron bound,
they give rise to different phenomenological scenarios. We will come back to this point again later.

But this is not the end of the story regarding the consequences of an early period of matter domination.
As it is well known, an early period of matter domination, triggered by a super heavy unstable but longlived particle which goes 
out of equilibrium at early times and comes to dominate the energy density of the Universe, leads to a reduction of the required number 
of e-folds before the end of inflation at which the scales of interest today left the horizon.
This reduction, which relaxes the flatness condition for the inflationary potential,  is due to the fact that the comoving horizon scale grows as $a^{1/2}$ 
during a matter dominated epoch in contrast to the radiation dominated one where the comoving horizon grows as $a$. As a consequence, 
the longer the period of matter domination, the smaller the growth of the universe from the end of inflation up today and therefore the smaller
the number of efolds required.   This reduction is given by \cite{Barenboim:2008zk}
\beq
\Delta N = \frac{1}{4}\log\left(\frac{a_{\rm end}}{a_{\rm start}}\right) \ , 
\eeq
where $a_{\rm end}$ and $a_{\rm start}$ are the scale factor at the beginning and end of the matter dominated era respectively. 
In terms of the parameters which define our model, \ie the mass and decay width of $X$, this ratio between the scale factors reads
\beq
\frac{a_{\rm end}}{a_{\rm start}} = \left(\frac{H_{\rm start}}{H_{\rm end}}\right)^{2/3}\approx 3.9 \left(\frac{g_*}{\Gamma_\X M_{Pl}}  (r M_\X)^4 \right)^{1/3}\;. 
\eeq
For values of the decay width close to its lower bound and masses of the order of $10^{15}$ GeV, this reduction turns out to be over 10 e-foldings.

Likewise, the end of a  matter dominated Universe driven by the decay of a long lived massive particle leads to an  important entropy production 
\beq
\frac{S_{\mbox{end}}}{S_{\mbox{start}}} = \left(\frac{T_{\rm end}a_{\rm end}}{T_{\rm start}a_{\rm start}}\right)^3 = \left(\frac{\left.\frac{a_{\rm end}}{a_{\rm start}}\right|_{\rm MD}}{\left.\frac{a_{\rm end}}{a_{\rm start}}\right|_{\rm RD}}\right)^3\simeq
12.2 \left(\frac{g_*}{\Gamma_\X^2 M_{Pl}^2}\right)^{\frac{1}{4}} \left(r M_\X \right)\,.
\eeq
As it is well known, supersymmetry as well as most of the theories beyond the Standard Model
 are riddled with new particles associated to new (and higher) energy scales which produce undesirable relics whose abundances, or they mere 
presence at certain times, 
do not agree with the current experimental observations of our universe, 
\eg moduli  and gravitinos. So a 
large release of entropy might help to dilute them, softening (or completely erasing) the constraints on their masses. Consequently, this 
scenario provides the same services as thermal inflation (regarding the unwanted relics) but without 
introducing another scalar particle into the theory\cite{Lyth:1995ka}. 
  
On the other hand, it is also important to note that the entropy production that can so nicely solve the unwanted relic problem, can also erase the baryon asymmetry 
produced at the electroweak scale. Such erasement is given by 
\beq
\eta = \eta_{\rm EW}\left(\frac{S_{\rm end}}{S_{EW}}\right)\;,
\eeq
where $\eta_{\rm EW} \approx n_B/s$  is the baryon to photon ratio produced at the electroweak scale and the entropy is given by $S= g_* a^3 T^3$.

As mentioned before, at late times into the matter dominated period $a \propto 1/ T^{8/3} $ and then
\bea
\eta = \eta_{\rm EW} \left( \frac{T_{\rm EW}}{T_{end}} \right)^5\;,
\eea
which in terms of the model reads
\bea
\eta = \eta_{\rm EW} \frac{1.5 \times 10^{42}}{g_*^{5/4}} \left(\frac{\Gamma_X}{T_{EW}}\right)^{5/2}\;.
\eea
It is thus clear that we need to generate a large baryon to photon ratio, a ratio of order one or even larger. This needs that the mechanism
for baryogenesis to  be orders of magnitude more efficient that the standard case, something clearly difficult but not impossible.

One can also see that the entropy production is directly proportional to the decay width, the larger the decay width, the less restrictive
the erasement becomes, one would be tempted then to push  into the large $\Gamma_X$ regime. However, large decay widths lead us to scenario (ii)
where the temperature is essentially given by the radiation coming from $X$ decays,
a scenario where the relaxation of the sphaleron bound is inversely
proportional to the decay width. So any gain in the relaxation of the sphaleron bound means a loss in the asymmetry produced. 
This tension between both scenarios may be seen explicitly in the tables of figure \ref{tb:M125}. Consequently it is clear that the ``optimal" case, where we maximize the relaxation of the sphaleron
bound and at the same time minimize the dilution of the asymmetry occurs when 
\bea
\rho_r^{\mbox{\small{old}}}  \bigm\vert_{\mbox{\small{EW}}} 
\; \; \approx  \;\; \rho_r^{\mbox{\small{new}}}\bigm\vert_{\mbox{\small{EW}}}\;.
\eea

\FIGURE[t]{
\centering
\begin{tabular}{c}
\begin{tabular}{cc|c|c|c|c|c|l}
\cline{3-7}
& & \multicolumn{5}{c|}{$M_X = 10^{12}\ {\rm GeV}$} \\ \cline{3-7}
& & $\rho_{\rm X }^{\rm ew} \ ({\rm GeV}) $ & $\rho_{\rm r,old}^{\rm ew}\ ({\rm GeV})$ & $\rho_{\rm r,new}^{\rm ew}\ ({\rm GeV})$&$T_d\ ({\rm GeV})$&$\Delta\left(\phi_c/T_c\right)$ \\ \cline{1-7}
\multicolumn{1}{|c}{\multirow{8}{*}{$\Gamma_X \ ({\rm GeV})$}} &
\multicolumn{1}{|c|}{$10^{-15}$} &$3.47\cdot10^{13}$ &$ 1.61$&$	9.94\cdot10^9$ &$ 17.10$ &$	0.10$     \\ \cline{2-7}
\multicolumn{1}{|c}{}                        &
\multicolumn{1}{|c|}{$10^{-17}$} &$3.48\cdot10^{17}$&$	3.48\cdot10^5$&$ 9.94\cdot10^9$&	$1.71$&	$0.23$     \\ \cline{2-7}
\multicolumn{1}{|c}{}                        &
\multicolumn{1}{|c|}{$10^{-19}$} &$5.29\cdot10^{20}$&	$6.08\cdot10^9$&	$3.88\cdot10^9$&	$0.17$	& $0.33$     \\ \cline{2-7}
\multicolumn{1}{|c}{}                        &
\multicolumn{1}{|c|}{$10^{-20}$} &$7.37\cdot10^{20}$&	$9.46\cdot10^9$&	$4.58\cdot10^8$&	$0.05$	& $0.34$     \\ \cline{2-7}
\multicolumn{1}{|c}{}                        &
\multicolumn{1}{|c|}{$10^{-21}$} &$7.61\cdot10^{20}$&	$9.87\cdot10^9$&	$4.65\cdot10^7$&	$0.01$	& $0.34$    \\ \cline{2-7}
\multicolumn{1}{|c}{}                        &
\multicolumn{1}{|c|}{$10^{-22}$} &$7.64\cdot10^{20}$&	$9.93\cdot10^9$&	$4.66\cdot10^6$&	$5.41\cdot10^{-3}$ &$0.34$    \\ \cline{2-7}
\multicolumn{1}{|c}{}                        &
\multicolumn{1}{|c|}{$10^{-23}$} &$7.64\cdot10^{20}$&	$9.93\cdot10^9$&	$4.66\cdot10^5$ &	$1.71\cdot10^{-3}$&	$0.34$    \\ \cline{2-7}
\multicolumn{1}{|c}{}                        &
\multicolumn{1}{|c|}{$10^{-24}$} &$7.65\cdot10^{20}$&	$9.93\cdot10^9$&	$4.66\cdot10^4$&	$5.41\cdot10^{-4}$&	$0.34$    \\ \cline{1-7}
\end{tabular}
\\
\\
\\
\begin{tabular}{cc|c|c|c|c|c|l}
\cline{3-7}
& & \multicolumn{5}{c|}{$M_X = 10^{15}\ {\rm GeV}$} \\ \cline{3-7}
& & $\rho_{\rm X }^{\rm ew} \ ({\rm GeV}) $ & $\rho_{\rm r,old}^{\rm ew}\ ({\rm GeV})$ & $\rho_{\rm r,new}^{\rm ew}\ ({\rm GeV})$&$T_d\ ({\rm GeV})$&$\Delta\left(\phi_c/T_c\right)$ \\ \cline{1-7}
\multicolumn{1}{|c}{\multirow{8}{*}{$\Gamma_X \ ({\rm GeV})$}} &
\multicolumn{1}{|c|}{$10^{-15}$} &$3.48\cdot10^{13}$ &$ 1.61\cdot10^{-4}$&$ 9.94\cdot10^9$&	$17.11$&	$0.11$   \\ \cline{2-7}
\multicolumn{1}{|c}{}                        &
\multicolumn{1}{|c|}{$10^{-17}$} &$3.48\cdot10^{17}$&$34.76$&	$9.94\cdot10^9$&	$1.71$	& $0.23$     \\ \cline{2-7}
\multicolumn{1}{|c}{}                        &
\multicolumn{1}{|c|}{$10^{-19}$} &$3.48\cdot10^{21}$&	$7.49\cdot10^6$&	$9.94\cdot10^9$&	$0.17$	& $0.36$    \\ \cline{2-7}
\multicolumn{1}{|c}{}                        &
\multicolumn{1}{|c|}{$10^{-20}$} &$2.25\cdot10^{23}$&	$1.94\cdot10^9$&	$7.99\cdot10^9$&	$0.05$	& $0.42$    \\ \cline{2-7}
\multicolumn{1}{|c}{}                        &
\multicolumn{1}{|c|}{$10^{-21}$} &$6.81\cdot10^{23}$&	$8.52\cdot10^9$&	$1.39\cdot10^9$&	$0.02$	& $0.44$\\ \cline{2-7}
\multicolumn{1}{|c}{}                        &
\multicolumn{1}{|c|}{$10^{-22}$} &$7.58\cdot10^{23}$&	$9.82\cdot10^9$&	$1.47\cdot10^8$ &	$5.41\cdot10^{-3}$&	$0.44$\\ \cline{2-7}
\multicolumn{1}{|c}{}                        &
\multicolumn{1}{|c|}{$10^{-23}$} &$7.65\cdot10^{23}$&	$9.93\cdot10^9$&	$1.47\cdot10^7$&	$1.71\cdot10^{-3}$&	$0.44$\\ \cline{2-7}
\multicolumn{1}{|c}{}                        &
\multicolumn{1}{|c|}{$10^{-24}$} &$7.65\cdot10^{23}$&	$9.93\cdot10^9$&	$1.47\cdot10^6$&	$5.41\cdot10^{-4}$&	$0.44$ \\ \cline{1-7}
\end{tabular}
\end{tabular}
\caption{Value of the important parameters taking special role in our particular scenario dominated by a heavy particle with mass $M_X= 10^{12} \; {\rm and}  \; 10^{15} \; {\rm GeV}$ and for a set of different decay widths $\Gamma_X$. $T_d$ stands for the recovery point of the common radiation dominated era. We have assumed that the thermal history of the Universe begun at about $T\sim 10^{17} \; {\rm GeV}$.}
\label{tb:M125}
}

In this case, the sphaleron bound, for a broad range of $M_X$ values, weakens to 
\beq
\phitc\gtrsim [0.64-0.69]\;,
\eeq
 which may be sufficient to open the window to electroweak baryogenesis in many extensions of the SM and particularly in the MSSM. 
Figure \ref{fig:phitcmq} shows the ratio of the temperature dependent vev at the critical
temperature to the critical temperature for different  Higgs masses as a function of the stop mass (each pair Higgs-stop
mass determines the corresponding mixing). The shadowed region
signals the reduction that can be obtained for a range of masses and decay widths characterizing the longlived but unstable particle $X$
from the usual $\phi_c/T_c > 1$ bound for preservation of the asymmetry in the standard cosmological scenario. 
From there it can be clearly seen
that a Higgs on the 125-135 GeV range could be made compatible with electroweak baryogenesis, if the thermal history of our universe
includes a prolongued period of matter domination.

At this point, we must discuss if a particle exists with the characteristics described above. We are looking for a super heavy particle, 
with an extremely long lifetime in thermal equilibrium at temperatures above its mass. The only particle that appears in (almost) all the extensions 
of the Standard Model that fulfils these requirements is beyond any doubt the right handed neutrino. Right handed neutrinos 
through the see-saw mechanism are the
fine-tuning-free minimal extension of the Standard Model able to reproduce the only evidence we have observed so far beyond the
Standard Model, the light  neutrino masses (if their Yukawa couplings are small enough).

Of course there are not one but three right handed neutrinos, and their mass matrices and Yukawa couplings are strictly 
model dependent. However, in a fairly model independent way the mass and lifetime of our $X$ particle, if a right
handed neutrino, satisfies 
\bea
\Gamma \propto \frac{m_i M_i^2}{v^2}
\eea
being $m$ the observed light neutrino mass of flavour $i$, which implies  a hierarchycal
scenario with an negligible small lightest mass (indistinguishable from zero from an experimental
point of view).

\FIGURE[t]{
\centering
\includegraphics[scale=0.8]{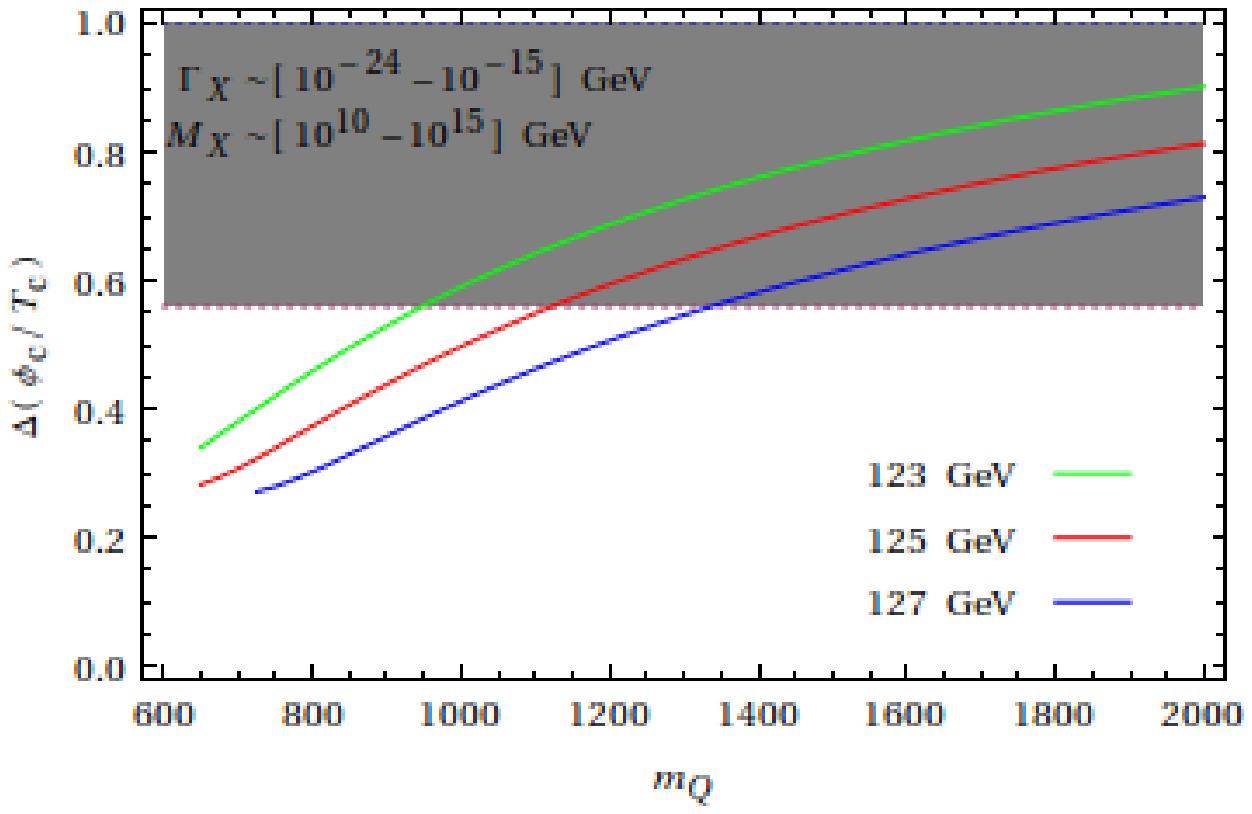}
\caption{Dependence on ratio of the temperature dependent vev to the temperature at $T_c$  on the stop mass ($\sim m_Q $)
for a fixed Higgs mass. The shadowed region shows the relaxation on the sphaleron
bound that can be obtained by an erly period of matter domination triggered by a long lived but unstable heavy particle}
\label{fig:phitcmq}
}

\section{First order phase transition in braneword cosmologies}

Alternatively to the previous scenario, a relaxation to the sphaleron bound can be obtained
by  modifying the underlying cosmology. In order to do so, we will  introduce ourselves into the braneworld language where we 
live in a brane embedded in a higher dimensional Universe. Within this scenario, one can consider different forms of the Stress-Energy 
momentum on the Bulk, which lead to the non standard behaviour of the Universe 
on the brane  we are looking for, by suitable choices of boundary conditions.

Regarding braneworlds, Randall-Sundrum argued that an ADS bulk and a brane with negative tension can provide a simple 
solution to the hierarchy problem \cite{Randall:1999vf}. Moreover, the Hubble rate on the brane under this scenario shows a non standard form
\beq
H^2 = \frac{8\pi}{3M_{Pl}^2}\rho\left(1 + \frac{\rho}{2\sigma}\right) + \frac{{\cal C}}{a^4}\;.
\eeq

On the other hand, Chung and Freese \cite{Chung:1999zs} showed that, in the context of braneworlds, it is possible to find 
any function of the FRW equation if one changes the stress energy tensor composition in the bulk. Therefore, in general, one can 
parametrize the expansion rate in the following form
\beq
H^2 = \kappa \rho + \mu \rho^n
\eeq
where $\kappa = \frac{8 \pi}{3M_{Pl}^2}$ and $\mu \sim {\cal O}({\rm GeV}^{-(4n-2)})$.
Notice that the geometry of such a Universe is flat and it is
trivial to see that each value of $n$ will lead to a different class of FRW equations. 

For $n < 2/3$, we find the so-called ``Cardassian models'' \cite{Freese:2002sq}, where one can explain the acceleration 
of a flat Universe at late times. 
In this work, however, we are interested in the opposite regime for $n$. We will show that, any $n$, with $n > 1$ can play an important role reopening the window for electroweak baryogensis 
without enlarging the particle content. In \cite{Servant:2001jh}, a study was done for a Randall-Sundrumm like Universe, which in the particular case of $n = 2$.
We will generalize this
analysis for a generic modified expansion rate, showing that the Randall-Sundrum is only one particular choice among all
the possible cases.

For simplicity, we will consider a radiation dominated Universe. There the expansion rate can be written as  
\beq
H^2 = \kappa \rho_r\left(1 + \frac{\rho_r^{n-1}}{M^{4(n-1)}}\right)\;,
\eeq
where $\rho_r$ is the radiation energy density and $M$ is the scale at which the transition to the usual FRW equation takes place. 
As explained before, contrary to the Cardassian models, we are seeking for 
departures of the standard expansion rate  at early times and therefore, we need
to explore  $n > 1$. Remember that $n<1$, provides late time accelerated expansion
and while it gives a nice explanation for a flat, expanding matter dominated universe,
it cannot play any role during the electroweak phase transition ($n=1$ recovers the
usual FRW).

The above expression may be rewritten in a straightforward manner as
\bea
H^2 =&& \kappa \rho_r(T)\left(1 + \frac{T^{4(n-1)}}{T_m^{4(n-1)}}\right) \nonumber\\
=&&\kappa \rho_r(T)\left[1 + \left(\frac{T}{T_m}\right)^{4(n-1)}\right]\;,
\eea
where $\rho_r(T)= \frac{\pi^2}{30}g_*T^4$ \, and $T_m$ is the matching temperature, the temperature
 at which we evolve from a Universe with a modified FRW constraint to the usual one.

\FIGURE[t]{
\centering
\includegraphics[scale=0.8]{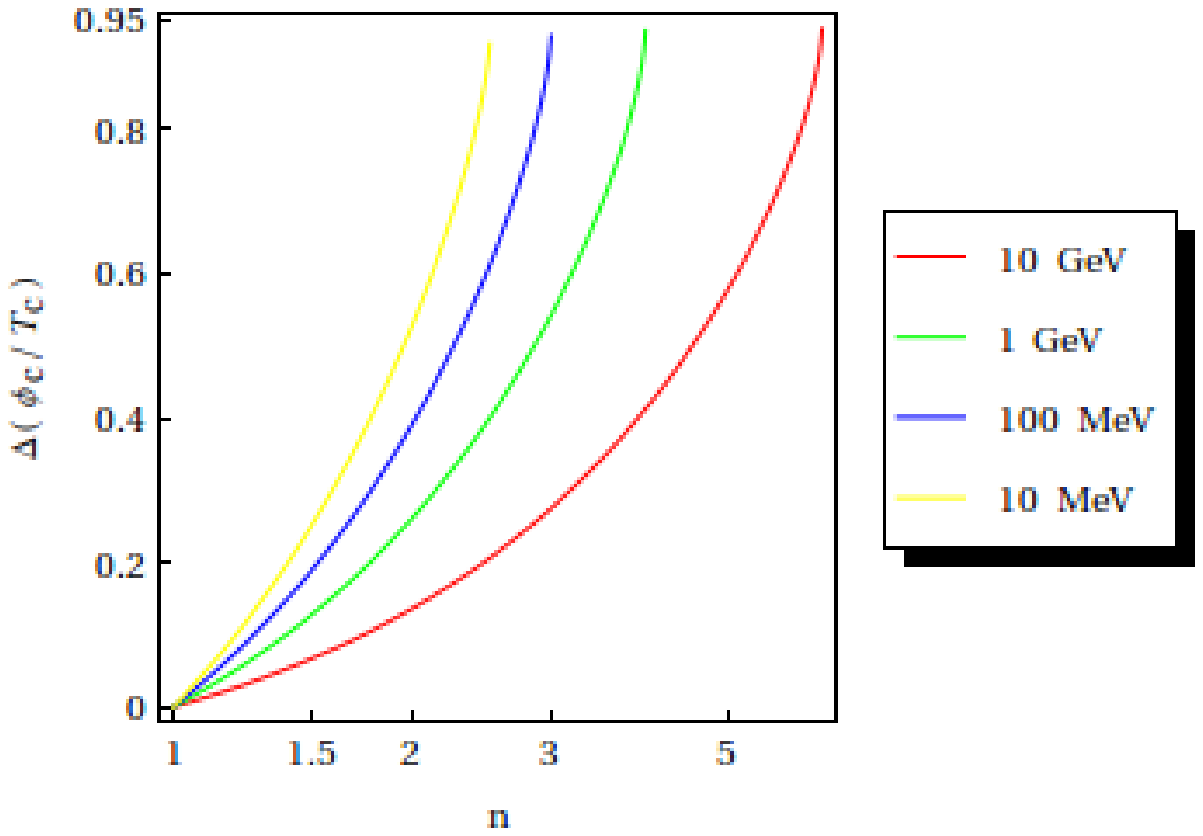}
\caption{Relative change on the sphaleron bound as a function of $n$ for different matching temperatures to the usual FRW scenario $T_m$ }
\label{fig:cardassian}
}

As we can see, at earlier epochs the second term dominates over the former one. Thus, using the value of the Hubble rate needed to make 
the phase  transition strongly first order, one can find a correlation between the matching temperature and the power of the cardassian model
\beq
H(T)= H_0(T)\left(\frac{T}{T_m}\right)^{2(n-1)}\;,
\eeq
where $H_0(T)=1.66g_*^{1/2} \frac{T^2}{M_{Pl}}$.

Using this expression for the Hubble rate at the electroweak scale, the sphaleron bound can change significantly for different values of $T_m$ and $n$. This is plotted in figure \ref{fig:cardassian}.

\section{Conclusions}

In this work we have shown that despite the fact that the available region in
parameter space for the SM and most of its extensions (most notably the MSSM)
for electroweak baryogenesis is highly constrained by experimental results,
it can be increased in some alternative scenarios, without enlarging its
particle content.

In particular we have discussed a scenario with an early period of matter
domination triggered by a long lived massive particle. In such a case, the
expansion rate can be orders of magnitude larger than the standard, radiation
dominated one and substantially relax the sphaleron bound.
The decay of this massive particle generates a huge entropy production that
can dilute away any unwanted relic, turning the constraints on  the
inflationary reheating temperature unnecessary.  In this respect we have shown
that an early period of matter domination mimics the nice effects of thermal inflation
with no additional particle content.
However this entropy
production also imposses strong constaints on the efficiency of the mechanism
for baryogenesis at the electroweak scale.

On the other hand, when analyzing thermal histories suffering very prolonged 
periods of matter domination
preceding the usual one,
one wonders whether there is any signature of their existence left that can be tested today.
An obvious place to look is of course, structure formation.

The total perturbation amplitude growth during  
the first matter-dominated phase will just be $a_{\mbox{end}}/a_{\mbox{start}}$.
Then, if the  
primordial perturbation amplitude (say, from inflation) is larger than  
$10^{-14}$, this just means that the structure becomes strongly non- 
linear for very prolonged periods of matter domination.

However we should keep in mind that the perturbation growth  only occurs for  
perturbations inside the Hubble radius, with the maximum growth  
occurring for those scales that came inside the Hubble radius at the  
beginning of the matter dominated epoch, \ie the smallest scales. Scales that  
entered the horizon later than  a time $t_i$  will grow by only $ a_{\mbox{end}}/a_i $,  
where $a_i$ is the scale factor at which they entered the horizon. 
Those will still be small physical scales today. Scales that  
never crossed inside the horizon during the early matter dominated epoch would not  
have this enhancement. So the prediction of this model for the  
perturbation power spectrum would be the ordinary LCDM + inflation  
spectrum on large scales with an enhancement of power that grows as a  
power of wavenumber$ k$ on small scales. The enhancement would set in 
gradually for$ k>k_e $, where  
$ k_e = (aH)_{\mbox{end}}$,  the comoving wavenumber above which  
the power spectrum is enhanced. Roughly, in the usual CDM model, the  
mass power spectrum $P \propto k^{n-4}$  on small scales, where $ n=0.96$ is  
the primordial spectral index from inflation. In this model with  
massive particle decay and an early matter dominated  epoch, the power spectrum on  
scales $ k > k_e $ will instead go as $ P \propto (k/k_e)^{n}$ , \ie the power  
grows on small scales and becomes non-linear on scales 
$ k > 10^{2.5}  k_e $.

However, the massive  
particle decays. And as we need the universe to become radiation dominated  
before BBN, the decay products should be relativistic. Relativistic  
particles will free-stream out of the mini-halos even if they are  
strongly non-linear in density contrast (their gravitational  
potentials are still weak). So the structures formed  will eventually evaporate,
leaving no trace of their existence behind.
Of course in scenarios more sophisticated than this simple one, there
will be traces of this first period of matter domination left. 
We will carry  out this study elsewhere.

We have also shown that a modification of the FRW equation can lead to
expansion rates at early times large enough to relax the sphaleron bound
to level consistent with current experimental bounds. In such scenarios 
the transition  to the standard cosmology takes place after the electroweak
phase transition and before BBN, not affecting then either structure formation
or the age of the universe.

\section*{ACKNOWLEDGMENTS}
It is a pleasure to thank Grigoris Panotopoulos, Mariano Quir\'os, Joe Lykken and specially Josh Frieman for discussions. The authors 
acknowledge financial support from spanish MEC and FEDER (EC) under grant 
FPA2011-23596,and Generalitat Valenciana under the grant PROMETEO/2008/004

\end{document}